# Collaborative Search Trails for Video Search

*Frank Hopfgartner     *†David Vallet     *Martin Halvey     *Joemon Jose

*Department of Computing Science,
University of Glasgow,
Glasgow, United Kingdom.

†Universidad Autónoma de Madrid,
Escuela Politécnica Superior Ciudad Universitaria de
Cantoblanco, 28049 Madrid, Spain.

{hopfgarf, halvey, jj} @ dcs.gla.ac.uk, david.vallet@uam.es

## ABSTRACT
In this paper we present an approach for supporting users in the difficult task of searching for video. We use collaborative feedback mined from the interactions of earlier users of a video search system to help users in their current search tasks. Our objective is to improve the quality of the results that users find, and in doing so also assist users to explore a large and complex information space. It is hoped that this will lead to them considering search options that they may not have considered otherwise. We performed a user centred evaluation. The results of our evaluation indicate that we achieved our goals, the performance of the users in finding relevant video clips was enhanced with our system; users were able to explore the collection of video clips more and users demonstrated a preference for our system that provided recommendations.

## Categories and Subject Descriptors
H.5.1 Multimedia Information Systems, H.5.3 Group and Organization Interfaces

## 1. INTRODUCTION
A number of technological developments have lead emergence of increased availability of video data. It is also now feasible to view video at home as easily as text-based pages were viewed when the Web first appeared. This has lead to a vast number of newspapers and television news broadcasts placing video online. In addition to this the improving capabilities and the decreasing prices of current hardware systems has lead to ever growing possibilities to store and manipulate videos in a digital format. Individuals now build their own digital libraries from materials created through digital cameras and camcorders, and use a number of systems to place this material on the web, as well as store them as their own personal collection. This has lead to an immediate and growing need for new retrieval methods, systems and techniques that can aid ordinary users in searching for and locating video scenes and shots that he/she requires from a vast ocean of video information. Current state of the art systems rely on using annotations provided by users, methods that use the low level features available in the videos or on an existing representation of concepts associated with the retrieval tasks. None of these methods are sufficient enough to overcome the problems associated with video search. Using annotations can provide problems, as users can have different perceptions about the same video and tag that video differently. Also users of video sharing systems do not provide sufficient annotations for retrieval. On the other hand, the difference between low-level data representation of videos and the higher level concepts users associate with video, commonly known as the semantic gap [2], provide difficulties for using these low level features. Bridging the semantic gap is one of the most challenging research issues in multimedia information retrieval today.

With the intention of overcoming some of the problems associated with video search we have developed a video retrieval system that uses the actions involved in previous searches to help and inform subsequent users of the system, through recommendations. This system does not require users to alter their normal searching behaviour, provide annotations or any other supplementary feedback. This is achieved by utilising the available information about user interactions. This system does not require a representation of the concepts in the video that the user wishes to retrieve, while still offering a workaround for the problems associated with the semantic gap [2]. We believe that the use of this system can result in a number of desirable outcomes for users. In particular, improved user performance in terms of task completion, it can aid user exploration of the collection and can also increase user satisfaction with their search and search results. An evaluative study was conducted, in order to examine and validate these assumptions. A baseline system that provides no recommendations was compared with our system that provides recommendations. The systems and their respective performances were evaluated both qualitatively and quantitatively.

## 2. SYSTEM DESCRIPTION
### 2.1 System
The interface for this system is shown in Figure 1 and can be divided into three main panels, search panel (A), result panel (B) and playback panel (C). The search panel (A) is where users formulate and carry out their searches. Users enter a text based query in the search panel (A) to begin their search. The users are presented with text-based recommendations for search queries that they can use to enhance their search (b). The users are also presented with recommendations of video shots that might match their search criteria (a), each recommendation is only presented once, but may be retrieved by the user at a later stage if they wish to do so. The result panel is where users can view the search results (B). This panel is divided into five tabs, the results for the current search, a list of results that the user has marked as relevant, a list of results that the user has marked as maybe being relevant, a list of results that the user has marked as irrelevant and a list of recommendations that the user has been presented with previously. Users can mark results in these tabs as being relevant or irrelevant by using a sliding bar (c). In the result panel additional information about each video shot can be retrieved. Hovering the mouse tip over a video keyframe will result in that keyframe being highlighted, along with neighbouring keyframes and any text associated with the highlighted keyframe (d). The playback panel (C) is for viewing video shots (g). As a video is playing it is possible to view the current keyframe for that shot (e), any text associated with that keyframe (f) and the neighbouring keyframes. Users can play, pause, stop and can navigate through the video as they can on a normal media player, and also make relevance judgements about the keyframe (h). Some of these tools in the interface allow users of the system to provide the explicit and implicit feedback, which is then used to provide recommendations to future users. Explicit feedback is given by users by marking video shots as being either relevant, maybe relevant or irrelevant (c, h). Implicit feedback is given by users playing a video (g), highlighting a video keyframe (d), navigating through video keyframes (e) and selecting a video keyframe (e).

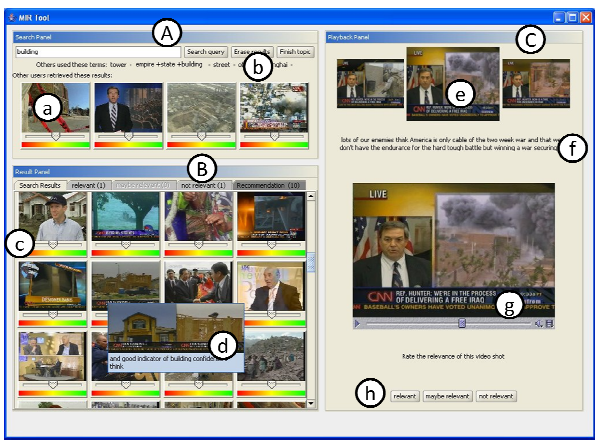

**Figure 1: Interface of the video retrieval system.**

In order to provide a comparison to our recommendation system, we also implemented a baseline system that provides no recommendations to users. The baseline system has previously been used for the interactive search task track at TRECVID 2006 [3]; the performance of this system was average when compared with other systems at TRECVID that year. Some additional retrieval and interface features were added to this system to improve its performance. Overall the only difference between the baseline and recommendation system is the provision of keyframe recommendations (a).

## 2.2 Graph Based Representation
For the implementation of our recommendation model based on user actions, there are two main desired properties of the model for action information storage. The first property is the representation of all of the user interactions with the system, including the search trails for each interaction. This allows us to fully exploit all of the interactions to provide richer recommendations. The second property is the aggregation of information from multiple sessions and users into a single representation, thus facilitating the analysis and exploitation of past information. To achieve these properties we opt for a graph-based representation of the users' interactions. We take the concept of trails from White et al [5]; however unlike White et al. we do not limit the possible recommended documents to those documents that are at the end of the search trail. The reason for this is that we believe that during an interactive search the documents that most of the users with similar interaction sequences interacted with are the documents that could be most relevant for recommendation, not just the final document in the search trail. Similar to Craswell and Szummer [1], our approach represents queries and documents in the same graph, however we represent the whole interaction sequence, unlike their approach, where the clicked documents are linked directly to the query node. This is because once again we want to recommend potentially important documents that are part of the interaction sequence. Another difference between our approach and previous work is that we take into consideration other types of implicit feedback actions, related to multimedia search, e.g. length of play time, browsing keyframes etc., as well as click through data. This additional data allows us to provide a richer representation of user actions and potentially better recommendations. In our system we recommend both queries and documents to the users, these recommendations are based on the status of the current user session. Full details of this representation and the recommendation algorithms are available in [4].

## 3. EXPERIMENTAL METHODOLOGY
In order to determine the effects of implicit feedback users were required to carry out a number of video search tasks based on the TRECVID 2006 evaluations [3]. For our evaluation we focused on search tasks from the interactive search track. For this evaluation we chose the four tasks for which the median precision in the 2006 TRECVID workshop was the worst. In essence these are the most difficult tasks. For our evaluation we adopted 2-searcher-by-2-topic Latin Square designs. Each participant carried out two tasks using the baseline system, and two tasks using the recommendation system. The order of system usage was varied as was the order of the tasks; this was to avoid any order effect associated with the tasks or with the systems. Each participant was given five minutes training on each system and each participant was allowed to carry out training tasks. These training tasks were the tasks for which participants had performed the best at TRECVID 2006. The users were given the evaluation topics and a maximum of fifteen minutes to find shots relevant to the topic. Although they were carrying out different tasks, the recommendations received were based on a single graph for the four tasks plus two training tasks. The users could carry out text based queries. For each participant their interaction with the system was logged, the videos they marked as relevant were stored and they also filled out a number of questionnaires at different stages of the experiment. The shots that were marked as relevant were then compared with the ground truth in the TRECVID collection.

## 4. RESULTS
24 participants took part in our evaluation. The participants were mostly postgraduate students and researchers at a university. The participants consisted of 18 males and 6 females with an average age of 25.2 years (median: 24.5) and an advanced proficiency with English. The participants indicated that they regularly interacted with and searched for multimedia. The participants were paid a sum of £10 for their participation in the experiment, which took approximately 2 hours. At the beginning of the evaluation there was no pool of implicit actions, therefore the first group of four users received no recommendations; their interactions formed the training set for the initial evaluations. The results of the user trials were analysed with respect to task performance, user exploration and user perceptions.

## 4.1 Task Performance

We begin our analysis by looking at the average interaction value that each video had been assigned based on the user interactions. These interaction values are a sum of the edge weights leading to a particular node. We wished to see if relevant documents did in fact receive more user interaction. The average interaction value was just 1.23, with irrelevant documents having an average value of 1.13 and relevant documents having an average of 2.94. This result shows that relevant documents receive more interactions from the users of the system. Up until a certain point as the interactions from previous users increases so does the probability of the document being relevant (see Figure 2). For some of the documents with higher relevance values the probability tails off slightly. There were two main reasons that a small number of irrelevant documents had high relevance values. Firstly, there were shots that seemed relevant at first glance but upon further investigation were not relevant; participants had to interact with the shot to investigate this. Secondly, there were a number of shots that appeared in the top of the most common queries, thus increasing the chances of participants interacting with those videos.

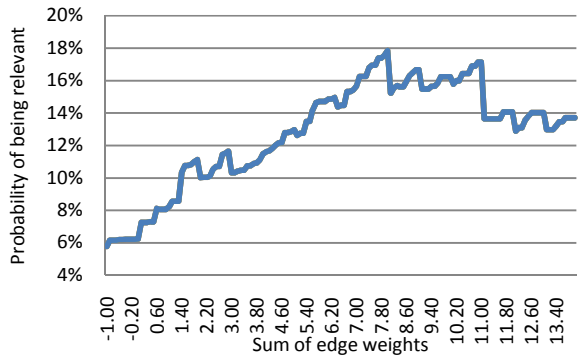

**Figure 2: Probability of a document being relevant given a certain level of interaction. The y-axis represents probability that the video is relevant and the x-axis represents assigned interaction value in our graph.**

Since we were using the TRECVID collection and tasks, we were able to calculate the P@N values for the both systems for varying values of N and also the mean average precision (MAP) for both systems for different groups of users. The results show that the system that uses recommendations outperforms the baseline system in terms of P@N. The shots returned by the recommendation system have a much higher precision over the first 5-30 shots than the baseline system. We verified that the difference between the two P@N values for values of N between 5 and 100 was statistically significant using a pair wise t-test (p = 0.0214, t = 3.3045). It was also found that the MAP of the shots that the participants selected using the recommendation system is higher than the MAP of the shots that the participants selected using the baseline system. We verified that the difference between the two sets of results were statistically significant using a pair wise t-test (p = 0.0028, t = 6.5623). The general trend is that the MAP of the shots found using the recommendation system is increasing with the amount of training data that is used to propagate the graph based model. While these results show that the users are seeing more accurate results and finding more accurate results, this is not telling the full story. In a number of scenarios users will just want to find just one result to satisfy their information need. Figure 3 shows the average time in seconds that it takes a user to find the first relevant shot for both the baseline and the recommender systems.

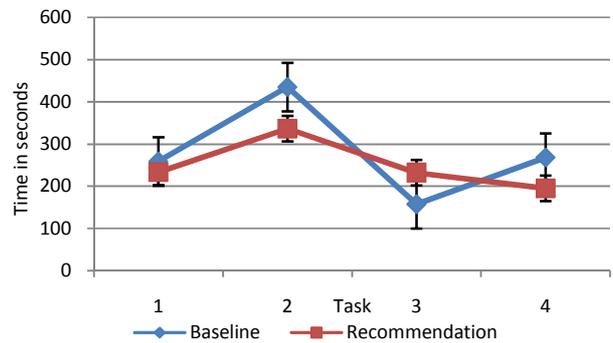

**Figure 3: Average time in seconds to find first relevant shot for Baseline and Recommendation Systems**

In Figure 3 it can be seen that for three of the four tasks the users using the baseline system find their first relevant result more quickly than the users using the baseline system. The one task for which the baseline system outperforms the recommendation system is due to the actions of two users who did not use the recommendations. However, a closer examination of the users who did use the recommendations found that three users found relevant shots in less than one minute, none of the users using the baseline system managed to find relevant shots in less than a minute. The variance in results for this particular task and system combination are reflected in the error bars for that point in Figure 3. Overall the difference in values is not statistically significant, but a definite trend can be seen. The results presented in this section have shown that users do achieve more accurate results using the system that provides recommendations.

## 4.2 User Exploration

We begin our investigation of user exploration by briefly analysing user interactions with the system. Table 1 outlines how many times each action available was used across the entire experimental group. It can be seen, in Table 1, that during the experiment the participants entered 1083 queries. Across the 24 users and 4 topics there is relatively little repetition of the exact same queries, there were 621 unique queries out of 1083 total queries. In fact only 4 queries occur 10 times or more, and they were all for the same task. This task had fewer facets than the others, and thus there was less scope for the users to use different search terms. This indicates that the participants took a number of different approaches and views on the tasks, indicating that their actions were not determined by carrying out the same tasks. The figures in Table 1 also show that participants play shots quite often. However, if a video shot is selected then it plays automatically in our system. This makes it more difficult to determine whether participants are playing the videos for additional information. To compensate for this we only count a play action if a video plays for more than 3 second. Another of the features that was most widely used in our system was the tooltip feature. The tooltip highlighting functionality allowed the users to view neighbouring keyframes and associated text when moving the mouse over one keyframe; this meant that the participants could get context and a feel for the shot without actually having to play that shot. This feature was used on average 42.3 (with a median of 38) times per participant per task when viewing a static shot. In contrast when participants could use this functionality while viewing a video, they choose not to. Instead they used other methods associated with browsing though a video. One of the preferred methods for browsing through a video adopted by some users was to skip through its shots, a sort of fast forward function.

| Action Type | Occ. | Action Type | Occ. |
| --- | --- | --- | --- |
| Query | 1083 | Play | 7598 |
| Mark Relevant | 1343 | Browse keyframes | 814 |
| Mark Maybe Relevant | 176 | Navigate within a video | 3794 |
| Mark Not Relevant | 922 | Tooltip | 4795 |
| View | 3034 | Total Actions | 23559 |

**Table 1: Action type and the number of occurrences during the experiment**

While these figures indicate how the users used the system, not all of these interactions were captured in the graph that provided recommendations. To continue our investigation of user exploration we analysed the graph of interactions. The number of nodes, the number of unique queries and the number of links that were present in the graph were analysed, at each stage where the graph had additional information for previous users added. It was found that the number of new interactions with the collection increases with the number of participants. The majority of nodes in our graph are video shots, as the number of participants increases so does the number of unique shots that have been viewed. On further investigation of the graph and logs it was found that, 49% of documents selected by users 1-12 were selected at least by one user in 13-24. Users 1-12 clicked 1050 unique documents, whereas users 13-24 clicked 596 unique documents. These results give an indication that further participants are not just using the recommendations to mark relevant videos, but also interacting with further shots. The results in this section indicate that users were able to explore the collection to a greater extent, and also discover aspects of the topic that they may not have considered. In order to validate this finding we must analyse the user perceptions of the tasks.

## 4.3 User Perceptions

In post search task questionnaires we solicited subjects' opinions on the videos that were returned by the system and also on the systems themselves. We wanted to discover if participants explored the video collection more based on the recommendations or if it in fact narrowed the focus in achievement of their tasks. From the results of the questionnaires the trend is that participants have a better perception of the video shots that they found during their tasks using the recommendation system. It also appears that the participants believe more strongly that this system changed their perception of the task and presented them with more options. We also found that the initial ideas that the participants had about relevant shots were dependent on the task ($p < 0.019$ for significance of task), where as the changes in their perceptions were more dependent on the system that they used rather than the task, as was the participants belief that they had found relevant shot through the searches ($p < 0.217$ for significance of system). After completing all of the tasks and having used both systems the participants were asked to complete an exit questionnaire where they were asked which system they preferred for particular aspects of the task, they could also indicate if they found no difference between the systems. The users were also given some space where they could provide any feedback on the system that they felt may be useful. It was found that the participants had a strong preference for the system that provided the recommendations. It is also encouraging that the participants found there to be no major difference in the effort and time required to learn how to use the recommendations that are provided by the system with recommendations.

## 5. CONCLUSIONS

There are a number of conclusions that can be made about our approach for using community based feedback to provide recommendations. We have presented an approach and a system for using feedback from previous users to inform and aid users of a video search system. The recommendations provided are based on user actions and on the previous interaction pool. The use of this system resulted in a number of desirable outcomes for the users. The performance of users of the recommendation system in terms of task completion improved with the use of recommendations based on feedback. The users were able to explore the collection to a greater extent and find more aspects of the task. Finally the users had a definite preference for the recommendation system in comparison with the baseline system, and perceived no additional overhead in using the recommendation system. In conclusion, our results have highlighted the promise of our implementation for using a community of user actions to alleviate the major problems that users have while searching for multimedia, thus presenting a potentially important step towards bridging the semantic gap [2].

## 6. ACKNOWLEDGEMENTS

This research work was partially supported by the European Commission under contracts: K-Space (FP6-027026) and SEMEDIA (FP6-045032).